\documentclass[aps,epsfig,twocolumn,showpacs,preprintnumbers,amsmath,amssymb]{revtex4}

\usepackage{graphicx}
\usepackage{dcolumn}
\usepackage{bm}

\begin{document}

\preprint{}

\title{Comparison of s- and d-wave gap symmetry on nonequilibrium
superconductivity}
\author{E.J. Nicol}
\affiliation{Department of Physics, University of Guelph,
Guelph, Ontario, N1G 2W1, Canada}
\author{J.P. Carbotte}
\affiliation{Department of Physics and Astronomy, McMaster University,
Hamilton, Ontario, L8S 4M1, Canada}
\date{\today}

\begin{abstract}
Recent application of ultrafast pump/probe optical techniques to
superconductors has renewed interest in nonequilibrium superconductivity
 and the predictions
that would be available for novel superconductors, such as the high 
$T_c$ cuprates. 
We have re-examined two of the classical models 
which have been used in the past to interpret nonequilibrium experiments 
with some success: the $\mu^*$ model of Owen and Scalapino
and the $T^*$ model of Parker. Predictions depend on
pairing symmetry. 
 For instance, the gap suppression due to
the excess quasiparticle
density $n$ in the $\mu^*$ model, varies as
$n^{3/2}$ in d-wave as opposed to $n$ for s-wave. 
Finally, we consider these models in the context of SIN tunneling and 
optical excitation experiments. While we confirm that recent
pump/probe experiments in YBCO, as presently interpreted, are in
conflict with d-wave pairing, we refute the further claim that they agree
with s-wave.
\end{abstract}

\maketitle

\section{Introduction}

The field of nonequilibrium superconductivity was very active
 throughout the late 1970's to mid 1980's when
it was realized that novel effects in the superconducting state 
could be induced by converting the electron distribution function
into a nonequilibrium one.\cite{langenberg} Different experimental
techniques were used to prepare such a nonequilibrium state,
for example, tunnel injection and optical irradiation, and
a body of work arose from both experimental and theoretical
efforts in this area. 
A useful summary of this work near the
end of this period of time can be found in a book edited
by Langenberg and Larkin\cite{langenberg} and other
broad-based texts have also appeared more recently\cite{kopnin,gulian}.

The advent of high $T_c$ cuprate superconductivity in 1986 interrupted 
work in this and other areas as the community turned its attention to this
new challenge and, consequently, extensive work in the
area of nonequilibrium superconductivity has
languished until more recently.
However, during the period following the original burst of activity,
new state-of-the-art experimental probes have been developed which
provide excellent opportunities for renewed interest in this field,
not to mention the potential for new insights provided by the new
generation of materials exhibiting novel superconductivity, such as
the cuprates. Some of these probes which can be turned to this problem
are: STM, ultrafast lasers, spin-polarized tunneling injection, 
terahertz spectroscopy, etc.

As early as the mid-eighties, the pump/probe femtosecond spectroscopy
was exhibiting its potential as a technique for investigating nonequilibrium
phenomena in metals and superconductors.
In these experiments, an 
ultrafast laser pulse ($\sim 100$ fs) 
incident on a sample as a high energy ``pump''  quickly
excites the electrons out of equilibrium which
then relax back to thermal equilibrium with
the lattice via the electron-phonon interaction. 
 Another laser pulse delayed in time ``probes''
the system of electrons by reflection or transmission
spectroscopy. As the system of electrons relaxes,
 the transient reflectivity or transmissivity
decays with time over a scale of picoseconds or less
allowing this experiment to
probe carrier dynamics in a time-resolved fashion.
A theory was proposed by P.B. Allen\cite{allen}
 for the relaxation of
quasiparticles in the normal state, which could be measured in
these experiments, resulting in the extraction of the 
electron-phonon renormalization parameter $\lambda$ (as the quasiparticles
relax through interactions with the system of phonons).
Experiments were performed which measured this parameter
using Allen's theory and excellent agreement was found with other
values in the literature for both ordinary metals and 
superconductors in the normal state.\cite{brorson}
Indeed, this parameter was measured for the first time in Cr by this
technique.\cite{brorson} This extraordinary success has led experimentalists
to use the femtosecond laser as a probe of 
high temperature 
superconductivity\cite{hanothers,otherlaserexpts,kabanov,mihailovic,oxford}
 and in general
several groups have been developing ultrafast techniques of similar
sort for measuring nonequilibrium phenomena in 
superconductors\cite{carr,feenstra}.

Here, we are interested in the state that arises when the
nonequilibrium excitations, created by a laser
pulse or by tunneling injection, have fallen to the gap edge but have not
yet recombined into the condensate (bottleneck effect).
In the first case, there is some
debate amongst experimentalists as to whether
the high energy laser used for pumping and probing can truly
measure the distribution of quasiparticles at low energy and
several groups are developing techniques to probe at lower energy
of order of the gap to address this issue.

The main thrust of our work 
has involved the use of two models employed in the past to describe
a nonequilibrium distribution of quasiparticles:
the $T^*$ model of Parker\cite{parker} which
uses an equilibrium distribution function
at an effective temperature $T^*$ relative to the bath temperature $T$ and
the  $\mu^*$ model, originally proposed by Owen and Scalapino\cite{owen},
 where the system is described in terms of
a new chemical potential
for the excited quasiparticles. 
The former approach 
has been used by Kabanov {\it et al.}\cite{kabanov} to analyze their optical
data, whereas the latter approach has been used for 
 systems where excess particles 
are injected into tunnel junctions\cite{ParkerWilliams}. 
While these two models  are somewhat simplified,
they appear to have been
 effective in capturing some of the experimental results
on low $T_c$ superconductors.

In section II, we calculate how the superconductivity is modified as a
function of the nonequilibrium excess quasiparticle 
number density $n$. This leads  to 
modifications in the gap which we calculate numerically for various
values of temperature  $T$ characterizing the sample before
irradiation as a function of $n$ in both $\mu^*$ and $T^*$ models
and for s- and d-wave. For $T= 0$ 
and in the limit of $n\to 0$, we also
obtain analytic results for the
gap reduction versus $n$, for the chemical potential in 
the $\mu^*$ model and for the nonequilibrium effective temperature for
the $T^*$ model, as well as for the free energy difference
between the nonequilibrium superconducting state and the corresponding
equilibrium normal state. The analytic limits are tested
against the numerical work and found to be close to the exact
results even as $n$ increases towards its critical
value where superconductivity is destroyed.
Results for d-wave are
compared with s-wave and important differences are established.
In section III,
as an explicit example of an application
of our results, we consider a S-I-N tunneling
junction with a nonequilibrium state on the superconducting
side which we assume can be described by a $\mu^*$ model. We
show that the current voltage characteristics
are modified in two major ways.
First the amplitude of the gap is reduced  because of the presence of a 
nonequilibrium number of excess quasiparticles $n$ and secondly
the entire characteristic is shifted upward by a factor of $n$ in
appropriate units. Also the voltage at which the current 
is zero can be used to measure
the chemical potential $\mu^*$. Separate measurements of the gap
reduction, the chemical potential, and the upward shift in I-V
characteristic would allow a consistency test of the model.
In section IV, we consider the specific case of
pump/probe experiments and agree with 
previous theoretical work\cite{kabanov}
that the existing data, as currently interpreted, is not consistent with 
d-wave gap symmetry, but disagree that it is consistent
with s-wave. In a final section V, we draw conclusions
and give a summary of our results.

\section{Theory}

We consider two models used in the past for the 
treatment of non-equilibrium superconductivity. For an s-wave BCS 
superconductor, Owen and Scalapino considered a state in which there exists
a finite distribution of excess quasiparticles at the gap energy in
addition to a condensate.
In their $\mu^*$ model,\cite{owen}
thermal equilibrium is assumed although chemical equilibrium is not
 for the paired and unpaired electrons. This is mimicked through
the introduction of a chemical potential $\mu^*$ in the Fermi function
which represents a constraint on the quasiparticle excitation number.
With this chemical potential the Fermi function is:
\begin{equation}
f(E_k-\mu^*)=[1+{\rm exp}\beta(E_k-\mu^*)]^{-1}
\end{equation}
with the BCS gap equation modified to be:
\begin{equation}
\frac{1}{N(0)V}=\int^{\omega_c}_0 \frac{d\epsilon_k}
{\sqrt{\epsilon_k^2+\Delta^2(n)}} \tanh(\beta(E_k-\mu^*)/2)
\end{equation}
where $V$ is the pairing potential, $N(0)$ is the electronic 
density of states at the Fermi surface in the normal state
and the excess quasiparticle density $n$ is given as:
\begin{equation}
n = \frac{1}{\Delta(0)}\int^\infty_0 [f(E_k-\mu^*)-f(E_k)]d\epsilon_k
\end{equation}
where $\beta = 1/(k_BT)$, $E_k = \sqrt{\epsilon_k^2+\Delta^2(n)}$,
and $k_B$ is the Boltzmann constant.
Here $n$ is measured in units of $4N(0)\Delta(0)$. The $4$ is 
introduced for spin and for particle-hole parts
of the excitation spectrum. 
$\Delta(0)\equiv\Delta(n=0)$ is the superconducting gap in the
equilibrium state, finite and isotropic over the entire Fermi
 surface for s-wave 
gap symmetry.This model will be applied later to discuss tunneling.

Alternatively, Parker\cite{parker}
 considered a $T^*$ model
where instead of a $\mu^*$ in the Fermi function, a $T^*$ is used:
\begin{equation}
f(E_k,T^*)=[1+{\rm exp}(E_k/k_BT^*)]^{-1}
\end{equation}
with the other equations modified accordingly. This model is the one
used by Kabanov {\it et al.}\cite{kabanov}
in their analysis of the pump/probe data.

We consider first, the $\mu^*$ model for
 an s-wave BCS 
superconductor.
At zero temperature the existence of the excess
quasiparticles perturb the condensate by blocking
states which would otherwise be available to form the condensate in a 
variational sense, and this lowers the value of the gap.
The exact gap equation and 
relationship between chemical potential and $n$ are respectively,
\begin{equation}
\frac{\Delta(n)}{\Delta(0)}=\Biggl(\frac{\mu^*}{\Delta(0)}+n\Biggr)^2
\quad\hbox{\rm and}\quad n\Delta(0) = \sqrt{\mu^{*2}-\Delta^2(n)} 
\end{equation}
The first equation in (5) comes directly from the gap equation (2)
evaluated at zero temperature with reference made to the equilibrium
case which allows us to eliminate the pairing potential in
favour of $\Delta(0)$. The second follows from Eqn.~(3).
The grand potential $\Omega^S(n)$
(the familiar formula is given later for the anisotropic case in
Eqn.~(13)) in the isotropic case 
(at $T=0$) is
\begin{equation}
 \frac{\Delta\Omega(n)}{N(0)}
\equiv \frac{\Omega^S(n)-\Omega^N(0)}{N(0)}
=-\frac{1}{2}\Delta^2(n)-2\mu^*\sqrt{\mu^{*2}-\Delta^2(n)},
\end{equation}
where this is the
difference between the nonequilibrium superconducting
state and its normal equilibrium counterpart ({\it i.e.} with no
excess quasiparticles).
The difference normalized to  the equilibrium superconducting state 
condensation energy is
\begin{equation}
\frac{2\Delta\Omega(n)}{N(0)\Delta^2(0)}=
\frac{2[\Omega^S(n)-\Omega^N(n=0)]}{N(0)\Delta^2(0)}
\approx -1+8 n^2
\end{equation}
to lowest order in $n$.
To obtain Eqn.~(7) we have used expressions for $\Delta(n)/\Delta(0)$
and for $\mu^*/\Delta(0)$ valid to second order in $n$.
They are $\Delta(n)/\Delta(0)=1-2n-2n^2$
and $\mu^*/\Delta(0)=1-2n-3n^2/2$ (entered in Table I to lowest order).  
If we add to the grand potential, $\Delta\Omega(n)$, the number of
excess quasiparticles multiplied by the chemical potential, {\it i.e.} 
 $\mu^*\bar n$ where $\bar n$ is the first term of Eqn.~(3),
normalized in the same way
as Eqn.~(7), we get the normalized free energy difference at zero
temperature which we denote by $2\Delta F(n)/N(0)\Delta^2(0)$.
This is evaluated to be 
\begin{equation}
\frac{2\Delta F(n)}{N(0)\Delta^2(0)} \simeq -1+8n 
\end{equation}
(entered in Table I).

\begin{table}
\caption{Analytical forms for $n\to 0$ at $T=0$ in the $\mu^*$ model.
Note $n$ is in units of $4N(0)\Delta(0)$, where $N(0)$ is the single
spin density states and $\Delta(0)$ is the $T=0$ and $n=0$ gap (maximum
in d-wave). }
\begin{tabular}{ccc} 
$\mu^*$ model &s-wave & d-wave\\  
 \hline 
 $\Delta(n)/\Delta(0)$&$1-2n$&$1-\frac{4\sqrt{2}}{3}n^{3/2}$\\
 $2\Delta F(n)/N(0)\Delta^2(0)$& \qquad $-1+8n$ \qquad\qquad &$-\frac{1}{2}+\frac{16\sqrt{2}}{3}n^{3/2}$\\
 $\mu^*/\Delta(0)$&$1-2n$&$\sqrt{2}n^{1/2}$\\
\end{tabular}
\label{table1}
\end{table}

\begin{figure}[h]
\begin{picture}(250,200)
\leavevmode\centering\includegraphics{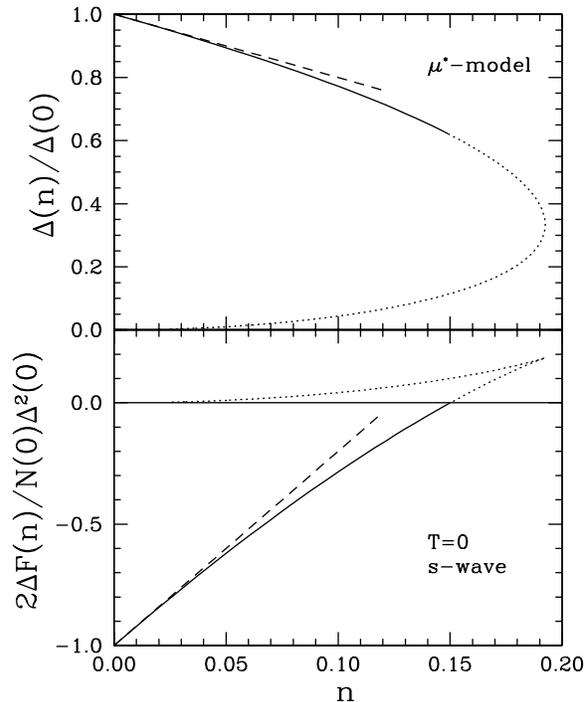}
\end{picture}
\vskip 60pt
\caption{Top frame: $\Delta(n)/\Delta(0)$ versus $n$, at 
$T=0$ for the $\mu^*$ model with an s-wave gap. 
The solid curve is
physical, the dotted curve is not. This latter curve
represents the case where the free energy of the normal
state is lower than that of the superconducting state as
shown in the bottom frame.
The presence of excess quasiparticles suppresses the gap
and eventually leads to a first order transition to the
normal state at $n=0.15$.
Bottom frame: $\Delta F = F_N-F_S$, the free energy difference, versus $n$.
 In both frames, the dashed curve is the small n limit
(see Table I). }
\label{fig1}
\end{figure}

In the top frame of Fig.~1, we present
our numerical results for the ratio $\Delta(n)/\Delta(0)$ as a 
function of excess quasiparticles 
$n$ (solid curve) and compare with the approximate
result $\Delta(n)/\Delta(0)=1-2n$ (dashed curve). We see
excellent agreement at small $n$.
As $n$ is increased, the 
continuation of the solid curve is denoted by the dots. It is
terminated at the point where the free energy for the 
nonequilibrium state becomes equal to its normal state value and
a first order transition occurs. This can
be seen more clearly in the bottom frame which shows the normalized
free energy difference of the nonequilibrium $(n\ne 0)$ state,
 $2\Delta F(n)/N(0)\Delta^2(0)$ 
as a function of $n$. The solid curve applies to the exact result
at $T=0$ while the dashed is the approximate result (Eqn.~(7).
 which fits
well the exact result at small $n$ and is semiquantitative in
the entire physical region. The first order phase transition to the
normal state occurs at $n_c\sim 0.15$.
The continuation of the solid line for the free energy difference to 
values of excess quasiparticles $n$ beyond the critical value is
indicated by a dotted curve just as in the top frame for the gap.
We note that both the gap $\Delta(n)$ and the free energy difference
$\Delta F(n)$ as a function of $n$ fold back on themselves
beyond a certain value of $n$, but that the free energy remains
positive for the entire dotted region
{\it i.e.} the nonequilibrium state has higher free energy
than does the normal state ($\Delta F(n)>0$) in this
region. 

\begin{figure}[h]
\begin{picture}(250,200)
\leavevmode\centering\includegraphics{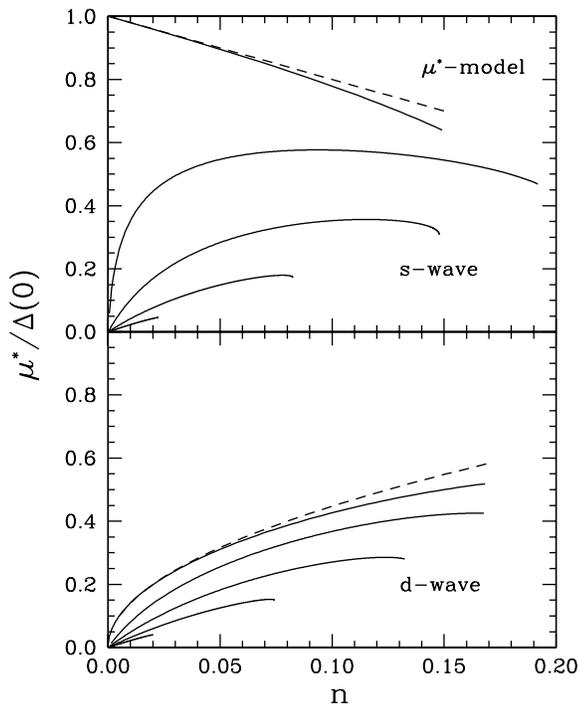}
\end{picture}
\vskip 60pt
\caption{The parameter $\mu^*$ versus $n$ for several temperatures
shown for the s-wave (top frame) and d-wave (bottom frame) gaps.
 The dashed curve is the small n limit (see Table I).
From top to bottom, the solid 
curves are for $T/T_c = t =0$, 0.3, 0.5, 0.7, 0.9.
Here only the physical part of the curves are shown.}
\label{fig2}
\end{figure}

Now we treat the d-wave case. The equation
relating $\mu^*$ to $n$ is $n\Delta(0) =\int^{\mu*}_0\bar N(E)dE$ where for
small $E$, $\bar N(E)\simeq E/\Delta(n)$. Here, $\Delta(n)$ 
is the maximum d-wave gap where the gap $\Delta(\phi)$ at any point $\phi$ 
(the polar angle for momentum) on the
2-dimensional Fermi circle in  the CuO$_2$ Brillouin zone is 
$\Delta(\phi)=\Delta(n)\cos(2\phi))$ 
with zeros in the ($\pi$,$\pi$)
direction and other symmetry related points. The small $n$ limit gives 
$\mu^*/\Delta(0)=\sqrt{2}n^{1/2}$ which differs radically from the 
s-wave case and reflects the gap symmetry with nodes
(see Table I). 
Numerical results for $\mu^*/\Delta(0)$ versus $n$ are
given in Fig.~2. 
The top frame applies to the s-wave case and is for
comparison with the bottom frame for d-wave. 
The dashed curves in both frames are
our approximate analytical results which are seen to match well the exact
results (solid curve for $T=0$) in the small $n$ limit. The
remaining curves are at finite temperature $T$ as indicated in
the caption, namely $T/T_c=0.3$, 0.5, 0.7 and 0.9. Several features
are worth noting. For s-wave,
the zero temperature behaviour of the chemical potential as a 
function of $n$ is qualitatively different from
the case for finite temperature. In the limit of $n\to 0$ {\it i.e.}
very few excess quasiparticles, the chemical potential must clearly be
equal to $\Delta(n=0)$ at $T=0$. In this case the lowest energy available
quasiparticle states are at $\Delta(0)$ where there is an inverse square
root singularity in the density of states and hence all the
excess quasiparticles can be accommodated at the gap energy.
As $n$ increases out of zero, the gap $\Delta(n)$ in the 
nonequilibrium state decreases from its value at $n=0$.
The inverse square root singularity shifts to lower energy
and there are now many states at and around $\Delta(n)$ and it turns out 
that all the excess quasiparticles can be accommodated in a small
energy range around the new gap value. We have already noted that
to second order in $n$,
$\Delta(n)/\Delta(0)$ and
$\mu^*/\Delta(0)$ differ by a factor
of $n^2/2$, specifically, $\mu^*/\Delta(0)=
[\Delta(n)/\Delta(0)] +n^2/2$ which implies that $\mu^*$
falls a few percent above the nonequilibrium
value of the gap in units of $\Delta(0)$. Note that the 
inequality $\Delta(n)/\Delta(0)
<\mu^*/\Delta(0)$ (at $T=0$ only), found to hold to second order in $n$,
was also verified in the numerical
work, which shows that the difference between $\Delta(n)$ and $\mu^*$
are always small even outside the validity of our expansion.
That this difference should be small is a reflection
of the square root singularity in the density of states.
 
The situation is very different in the d-wave case and in s-wave at finite
temperature. In these two cases the chemical potential becomes small
as $n\to 0$. For the d-wave case this is easily understood because
there is a small but finite density of states at any nonzero
value of energy $\omega\ne 0$.
The excess quasiparticles can occupy these states and hence $\mu^*\to 0$
as $n\to 0$. For the s-wave case at finite $T$ a different
argument holds. In this case the thermal factor $f(E_{\vec k}-\mu^*)$
gives the probability that the state $E_{\vec k}$ is occupied at finite $T$.
This probability can be increased over its value for $\mu^*=0$ simply
by having $\mu^*$ take on a small finite value
to accommodate the excess quasiparticles.
At low temperature, however, the thermal tails of the occupation
factor are small in the region of the gap and $\mu^*$ must increase fairly
rapidly as $n$ increases. This is seen most clearly
in the second highest curve in the
top frame of Fig.~2 which corresponds to $T/T_c=t=0.3$. 
Also as the temperature
is increased $\mu^*$ decreases as expected.
In the
d-wave case shown in the lower frame of Fig.~2, $\mu^*$
starts from zero at $n=0$ even at zero temperature because, 
as we have already indicated,
there are states available at any energy
above $\omega=0$. 
Comparing top and bottom frame we note that the chemical potential
for $t=0.3$ (to be specific) rises more rapidly in the s-wave
case and becomes bigger than for d-wave. This can be traced to the
fact that for d-wave the part of the density of states that is
occupied  by the excess quasiparticles is in the range 0 to $\mu^*$
while in the s-wave case it is the region just about the gap
$\Delta(n)$ which is relevant. As the temperature is increased towards
$T_c$, the differences in the quasiparticle density of states between
s- and d-wave become smaller and the chemical potentials start to
become very similar.
A second feature to be noticed is
that at finite $T$ the curves for $\mu^*$ extend to higher values of 
$n$ for the s-wave case than they do in the d-wave case although
the reverse is true at zero temperature. In all cases the
curves terminate when the free energy difference between normal and
nonequilibrium superconducting state becomes zero or there
are two solutions and the one with the lowest free energy is chosen. 
This occurs at
smaller values of $n$ for the d-wave case as compared with s-wave
for the given temperature $T \ne 0$ shown. 
We will return to this issue later on in our discussion of Fig. 4.

The gap
equation with a pairing potential of the form 
$V_{\bf kk'}=V\cos(2\phi')\cos(2\phi)$, where $\bf k$ 
is momentum on the Fermi surface, with a
distribution of excess quasiparticles included through the introduction
of a chemical potential takes the form
\begin{equation}
\frac{1}{N(0)V}=\Bigg\langle \int^{\omega_c}_0 \frac{2\cos^2(2\phi)d\epsilon_k}
{\sqrt{\epsilon_k^2+\Delta^2(n)\cos^2(2\phi)}} \tanh(\frac{\beta}{2}(E_k-\mu^*))\Bigg\rangle
\end{equation}
with $E_k=\sqrt{\epsilon_k^2+\Delta^2(n)\cos^2(2\phi)}$. The bracket
$\langle\cdots\rangle$ indicates the angular average and $\epsilon_k$
is energy integrated in a rim of width $\omega_c$ about the Fermi energy.
With reference to the
$n=0$ case ({\it i.e.}, $\mu^*=0$) we can rewrite Eqn.~(2) to read at $T=0$
\begin{equation}
\ln\Bigl(\frac{\Delta(n)}{\Delta(0)}\Bigr) = \Big\langle\int^{\omega_c}_0
\frac{-4\cos^2(2\phi)d\epsilon\,}{\sqrt{\epsilon^2+\Delta^2(n)\cos^2(2\phi)}}
\Big\rangle_{E_k\le\mu^*}
\end{equation}
where the integration over energy and
$\phi$ must duly take account of the restriction $E\le\mu^*$. For
small $n\to 0$ the leading order gives ($\Delta(n)=\Delta(0)+\delta\Delta(n)$)
\begin{eqnarray}
\frac{\delta\Delta(n)}{\Delta(0)}&=&-\frac{8}{\pi}\int^{\pi/2}_{\cos^{-1}(\mu^*/\Delta(n))} \cos^2\phi'
d\phi'\nonumber\\
&\times&\int^{\mu^*}_{|\Delta(n)\cos\phi'|}
\frac{dE}{\sqrt{E^2-\Delta^2(n)\cos^2\phi'}}
\end{eqnarray}
where we have changed from $\phi$ to $\phi'=2\phi$.
But the lower limit in the $\phi'$ integration in Eqn.~(11) restricts the 
integration to the nodal region which corresponds to $\phi'=\pi/2$.
We find
\begin{eqnarray}
\frac{\delta\Delta(n)}{\Delta(0)}&=&-\frac{8}{\pi}
\Bigl(\frac{\mu^*}{\Delta(n)}\Bigr)^{3}
\int^1_0 x^2 dx\ln\Big|\frac{1+\sqrt{1-x^2}}{x}\Bigr|\nonumber\\ 
&\simeq& -\frac{4\sqrt{2}}{3}n^{3/2}
\end{eqnarray}
(entered in Table I)
where we have used the relationship $\mu^*/\Delta(n)=\sqrt{2}n^{1/2}$
to lowest order.
In Fig.~3 we show exact numerical results for the normalized
gap $\Delta(n)/\Delta(0)$ as a function of $n$ for
the d-wave case (solid curve) and compare with our approximate result
(dashed curve) which applies only at small $n$. The agreement
is excellent even up to the point where the first
order transition to the normal state occurs. This is where the solid curve is
extended into the dotted curve. 
The gap function as a function  of $n$ is reduced less in d-wave (Fig.~3)
as compared to s-wave (Fig.~1) all the way to 
 $n=n_c$. 
The free energy difference
$\Delta F(n)$ becomes zero at $n=n_c\simeq 0.17$ which is to 
be compared with $\simeq 0.15$ in the s-wave case. At the critical $n$,
$\Delta(n)/\Delta(0)$ is almost 0.6 for s-wave while in the d-wave
case it has not yet reached 0.8. The blocking of states by the excess 
quasiparticles has much less effect on the condensate wavefunction
as reflected in the change in 
the value of the gap in d-wave than in s-wave because now
the excess quasiparticles accumulate in the nodal region.
Since the gap is zero or near zero in that region, it is clear that these
states do not contribute much to the lowering of energy brought about
by the formation of Cooper pairs. 

\begin{figure}[h]
\begin{picture}(250,200)
\leavevmode\centering\includegraphics{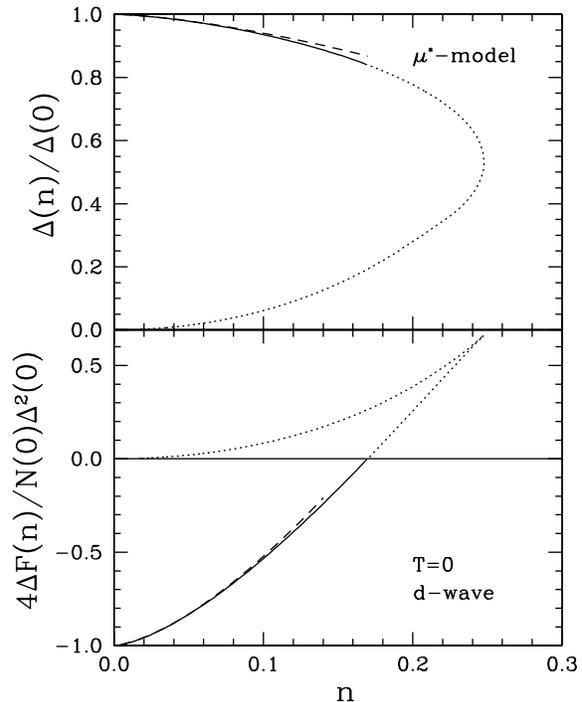}
\end{picture}
\vskip 60pt
\caption{The $\mu^*$ model at $T=0$ for a d-wave gap with the curves labelled
in the same manner as for Figure 1. The gap is suppressed
less rapidly in d-wave. The presence of excess quasiparticles, 
which normally weaken the condensate by blocking states, are less effective
in interfering  with the
formation of the superconducting wavefunction in d-wave as they
accumulate
at the nodes, in the first instance, which is a region where
the gap is close to zero.}
\label{fig3}
\end{figure}

To establish where this first order
transition occurs, we need the free energy.
The formula for the grand potential for the superconducting state
with $n$ excess quasiparticles is 
\begin{eqnarray}
\Omega^S(n) &=& 2k_BT\sum_k\ln(1-f(E_k-\mu^*))\nonumber\\
&+&\sum_k[\epsilon_k-E_k
+\frac{\Delta_k^2}{2E_k}(1-2f(E_k-\mu^*))]
\end{eqnarray}
and for the normal state with $n=0$ it is 
\begin{equation}
\Omega^N(0) = 2k_BT\sum_k\ln(1-f(|\epsilon_k|))+\sum_k(\epsilon_k-
|\epsilon_k|)
\end{equation}
The sum over $\bf k$ 
can be converted to energy
and the constant two dimensional electron density of states factor $N(0)$
taken out of the integration. In the limit $n\to 0$
\begin{eqnarray}
\frac{\Delta\Omega(n)}{N(0)}&\equiv& \frac{\Omega^S(n)-\Omega^N(0)}{N(0)}
= -\frac{1}{4}\Delta^2(n)\nonumber\\&+&4\int^{\mu^*}_0\bar N(E)(E-\mu^*)dE
- \frac{1}{2}I\Delta^2(n)
\end{eqnarray}
where $I$ is the same integral as appears on the right-hand side of
Eqn.~(11). The first term in (15) is the usual expression for the
condensation energy of a d-wave superconductor but with
$\Delta(n)=\Delta(0)(1-4\sqrt{2}n^{3/2}/3)$ replacing
the gap amplitude $\Delta(0)$ which applies to 
$n=0$. In $\Delta\Omega(n=0)/N(0)$ only
$\Delta^2(0)/4$ enters. The two extra terms
in Eqn.~(15) can be worked out analytically as $n\to 0$ and
lead to
\begin{equation}
\frac{\Delta\Omega(n)}{N(0)} = -\frac{1}{4}\Delta^2(n)
-\frac{2}{3}\frac{\mu^{*3}}{\Delta(n)}-\frac{1}{3}
\Bigl(\frac{\mu^*}{\Delta(n)}\Bigr)^3\Delta^2(n)
\end{equation}
only in the first term on the right-hand side of the equation
must we retain the $n$ dependence in $\Delta(n)$. 
The difference in grand potential $\Delta\Omega(n)$
 normalized to  $\Delta^2(0) N(0)/4$
is easily worked out to be
\begin{equation}
\frac{4\Delta\Omega(n)}{N(0)\Delta^2(0)}= 
-1-\frac{16\sqrt{2}}{3}n^{3/2}
\end{equation}
The normalized free energy $\Delta F$
is obtained 
by adding $\mu^* \bar n$ to Eqn.~(13) and after normalization
we get
\begin{equation} 
\frac{4\Delta F(n)}{N(0)\Delta^2(0)} = -1 +\frac{32\sqrt{2}}{3} n^{3/2}
\end{equation}
which is
entered in Table~I. Numerical results at any value of $n$ are shown
in the bottom frame of Fig.~3. The solid line is our
numerical result for $4\Delta F(n)/N(0)\Delta^2(0)$
 at $T=0$ and the dashed curve
our approximate result (Eqn.~(18)). The analytic
result agrees well with the full numerical solution at small $n$ and
differs slightly near the critical value of $n=n_c$ where the
first order transition to the normal state occurs at $n_c\approx 0.17$.

\begin{figure}[h]
\begin{picture}(250,200)
\leavevmode\centering\includegraphics{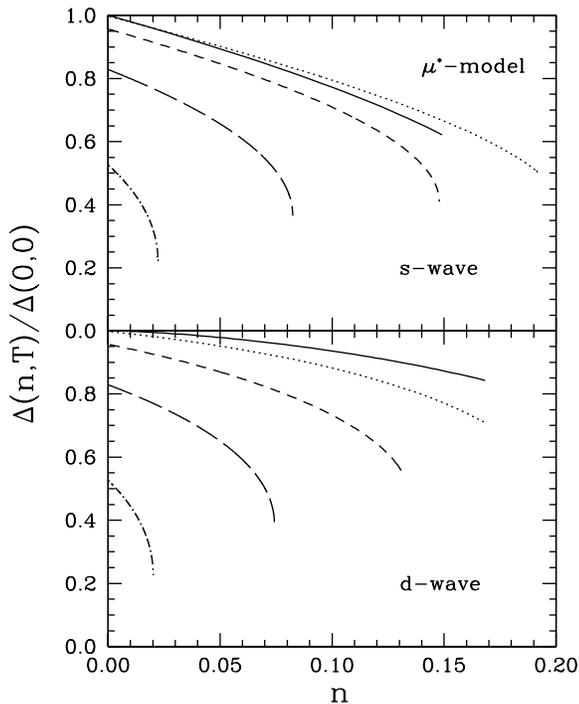}
\end{picture}
\vskip 60pt
\caption{The ratio of $\Delta(n,T)/\Delta(0,0)$ versus $n$ for
finite temperature in the $\mu^*$ model. The top frame is 
for the case of an s-wave gap and the bottom frame is for d-wave.
Curves are shown for
$T/T_c = t =0$ (solid curve), 0.3 (dotted), 0.5 (short-dashed),
0.7 (long-dashed), 0.9 (dot-dashed). Only the physical part of the
curves are shown.}
\label{fig4}
\end{figure}

In Fig.~4 we show our numerical results for the gap as a function of 
$n$ at various temperatures. 
Both frames are for the $\mu^*$ 
model.
The temperatures are $T/T_c=t=0$
(solid curve), 0.3 (dotted), 0.5 (short-dashed), 0.7 (long-dashed)
and 0.9 (dot-dashed). The top frame is for s-wave and is for comparison
with the bottom frame which is new and applies to d-wave.
Note that for s-wave, the $T=0$ curve is below the dotted curve for 
$t=0.3$. This agrees with findings  of Owen and Scalapino and has its
origin in the blocking process referred to previously. At zero
temperature the excess quasiparticles block important states which 
 cannot be used in the coherent superposition of states which form the Cooper
pair condensate. At finite temperature the blocking is less effective
because it is the states closest to zero energy that are the most
effective in forming the condensed pairs
while the thermal factor depopulates these
states. By contrast, for d-wave,
the $T=0$ curve is above the $t=0.3$ (dotted curve)
as we have already noted. In this instance
the blocking at $T=0$ is much less effective 
and consequently temperature is not as
important an effect.
We note again that, at $T=0$, the d-wave
gap is reduced less than in s-wave for the same value
of $n$ and that the critical
value of $n$, at which a first order transition
to the normal state takes place, is larger. 
At the higher temperatures shown, however,  the reverse holds.
Also, note that as the temperature rises towards $T_c$ the
difference between s- and d-wave get less pronounced as the
differences between the two
quasiparticle
density of states become small and also more states are involved.

The nonthermal quasiparticle distribution used in the $\mu^*$-model has
an interesting aspect in that it allows for the system to become unstable
to quasiparticle density fluctuations.\cite{chang74b,huberman}
 Essentially, if the
quasiparticles are injected uniformly in the sample, the
density fluctuations will act to draw off quasiparticles from
some regions thereby increasing the superconducting gap locally
and flowing those quasiparticles to other regions, causing an
accumulation which lowers the local gap, possibly even driving  the 
local region normal. This phase separation could be either a static or
a temporal structure. Such a state has been studied initially by Chang and
Scalapino\cite{chang74b} and Scalapino and Huberman\cite{huberman}
for the s-wave superconductor and experimental verification of
a density instability leading to an inhomogeneous multigap state has been
done by several groups\cite{inhomoexpts} using tunnel injection in thin
film nonequilibrium superconductors.
The theoretical signature of such an inhomogeneous
state in the $\mu^*$-model 
is that $\partial\mu^*/\partial n |_T <0$.\cite{chang74b,huberman} 
From Fig.~2, we find that the variation of $\mu^*$ with $n$
differs in s- and d-wave and by examining the slopes of these
curves, in particular, the point where the slope goes negative, we
can reproduce the s-wave phase diagram of Chang and Scalapino\cite{chang74b},
shown in the upper frame of Fig.~5, and provide the equivalent prediction
for d-wave in the bottom frame. The dashed curve in
these phase diagrams marks the boundary between the normal state (NS)
and the superconducting state (either homogeneous or inhomogeneous).
This boundary is entirely a first-order transition. The area
labelled IN, is the region of $n$ and $T$, where the slope of
the chemical potential curve is negative and an inhomogeneous  state
is predicted to exist. The solid line marks the boundary
between it and the homogeneous 
superconducting state (SC). There are qualitative differences between the 
s- and d-wave cases. The region of the inhomogeneous phase is quite large in
the s-wave case and almost non-existent in d-wave and at low temperature
the s-wave superconductor would likely be phase separated whereas, the
d-wave one would not be. While the inhomogeneous state may be of 
interest to study in itself, in the d-wave case it may be
encouraging to note that attempts at experimental verification of our
predictions for power law dependences, and other results presented in
this paper, are unlikely to be hampered by the presence of an 
inhomogeneous phase.

\begin{figure}[h]
\begin{picture}(250,200)
\leavevmode\centering\includegraphics{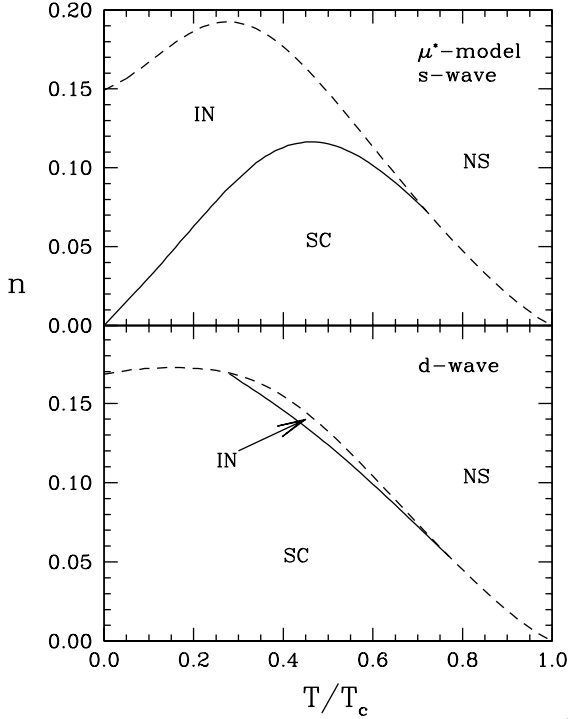}
\end{picture}
\vskip 60pt
\caption{The phase diagrams calculated in the 
the $\mu^*$ model for the s-wave (top)
and d-wave (bottom) gaps.
Based on the slope of $\mu^*$ versus $n$ one can
determine the region of the phase diagram where there
is a homogeneous (SC) and an inhomogeneous
superconducting state (IN). The
transition from the superconducting state to the normal
state (NS) is always first order and is represented by the dashed
line. }
\label{fig5}
\end{figure}

Next we consider briefly the case of the $T^*$ model
which is just a simple heating model if only the electronic system
is considered. 
Similar approximate analytic calculations can be done to get various
relationships in the limit $n\to 0$ for the case when the sample before 
irradiation is assumed to be zero.  These analytic derivations
are supplemented with full numerical work in which we also
consider the case when the sample is initially at finite temperature $T$.

We begin with the s-wave case and return to the gap equation shown
in Eqn.~(2), now modified according to Eqn.~(4) rather than Eqn.~(1). In the 
limit of $T\to 0$, the result for the lowest order correction to the gap
is well known\cite{fetter}:
\begin{equation}
\frac{\delta\Delta(n)}{\Delta(0)}=-\sqrt{\frac{2\pi k_BT^*}{\Delta(0)}}
e^{-\Delta(0)/k_BT^*}
\end{equation}
The relation between $n$ and $T^*$ can be trivially obtained as
$n=\sqrt{\pi T^*/2\Delta(0)}e^{-\Delta(0)/k_B T^*}$ and so
$\delta\Delta(n)/\Delta(0) = -2n$.

The d-wave case is not as well known and we include the critical steps here
\begin{eqnarray}
\ln\biggl(\frac{\Delta(n)}{\Delta(0)}\biggr) &=&
-4\int_0^{\pi/2}\frac{2d\phi'}{\pi} \cos^2\phi'\nonumber\\
&\times&\int^{\omega_c}_{\Delta(n)\cos\phi'} \frac{dE\, e^{-E/k_BT^*}}
{\sqrt{E^2-\Delta^2(n)\cos^2\phi'}}
\end{eqnarray}
which can be manipulated into
\begin{eqnarray}
\ln\biggl(\frac{\Delta(n)}{\Delta(0)}\biggr) &=&
-\frac{8}{\pi}\int_0^{\pi/2}d\phi \cos^2\phi'
e^{-\Delta(n)\cos\phi'/k_BT^*}\nonumber\\&\times&\int^{\omega_c}_0
\frac{dx\, e^{-x/k_BT^*}}{\sqrt{x(x+2\Delta(n)\cos\phi')}}
\end{eqnarray}
which can be manipulated into
\begin{eqnarray}
\ln\biggl(\frac{\Delta(n)}{\Delta(0)}\biggr) &=&
-\frac{8}{\pi}\int_0^{\pi/2}d\phi \cos^2\phi'
e^{-\Delta(n)\cos\phi'/k_BT^*}\nonumber\\&\times&\int^{\omega_c}_0
\frac{dx\, e^{-x/k_BT^*}}{\sqrt{x(x+2\Delta(n)\cos\phi')}}
\end{eqnarray}
The integral over $\phi'$ is peaked around $\cos\phi'=0$, {\it i.e.}
$\phi'$ near $\pi/2$ which allows us to approximate it by
\begin{eqnarray}
\ln\biggl(\frac{\Delta(n)}{\Delta(0)}\biggr) &=&
-\frac{8}{\pi}\int_0^{\infty}dy\, y^2
e^{-\Delta(n) y/k_BT^*}\nonumber\\&&\times\int^{\infty}_0
\frac{dx\, e^{-x/k_BT^*}}{\sqrt{x(x+2\Delta(n) y)}}
\end{eqnarray}
from which we get
\begin{equation}
\frac{\delta\Delta(n)}{\Delta(0)}=
-4\biggl(\frac{T^*}{\Delta(0)}\biggr)^3
\end{equation}
Also from the definition of $n$ we get immediately
\begin{equation}
n=\frac{\pi^2}{12}\biggl(\frac{T^*}{\Delta(0)}\biggr)^2
\end{equation}
Exact numerical results agree well with these approximate $n\to 0$ 
expressions which we summarize in Table~II.

\begin{table}
\caption{Analytical forms for $n\to 0$ at $T=0$ in the $T^*$ model.
Note $n$ is in units of $4N(0)\Delta(0)$, where $N(0)$ is the single
spin density states and $\Delta(0)$ is the $T=0$ and $n=0$ gap (maximum
in d-wave). }
\begin{tabular}{ccc} 
$T^*$ model &s-wave & d-wave\\  
 \hline 
 $\Delta(n)/\Delta(0)$&$1-2n$&$1-\frac{32}{\pi^3}(3n)^{3/2}$\\
 $T^*$ vs. $n$&  $n=0.94\sqrt{T^*/T_c}e^{-1.76T_c/T^*}$\qquad &$T^*/T_c=2.36n^{1/2}$\\
\end{tabular}
\label{table2}
\end{table}

\begin{figure}[h]
\begin{picture}(250,200)
\leavevmode\centering\includegraphics{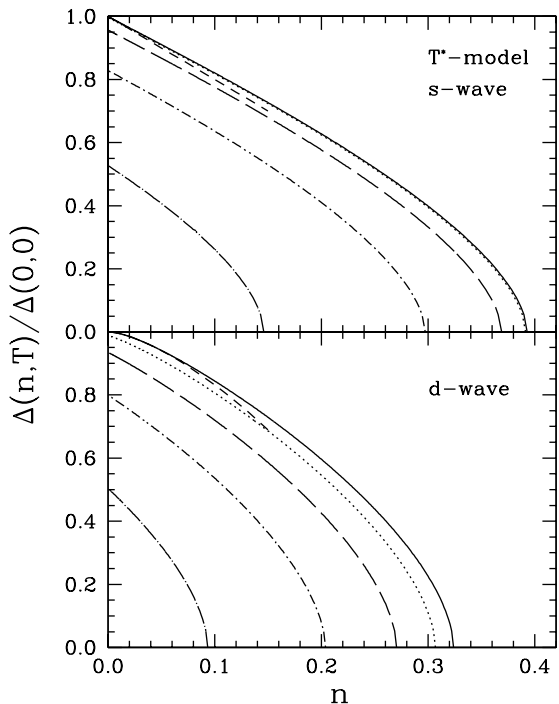}
\end{picture}
\vskip 60pt
\caption{The ratio of $\Delta(n,T)/\Delta(0,0)$ versus $n$ for
finite temperature in the $T^*$ model. The top frame is 
for an s-wave gap and the bottom frame is for d-wave.
Curves are shown for
$T/T_c = t =0.01$ (solid curve), 0.3 (dotted), 0.5 (long-dashed),
0.7 (dot-short-dashed), 0.9 (dot-long-dashed). The short-dashed 
curve is approximate
analytic form for low $n$ given in Table~II.}
\label{fig6}
\end{figure}

In Fig.~6 we show numerical results for 
$\Delta(n,T)/\Delta(0,0)$ versus $n$ where we have
normalized the maximum gap $\Delta(n,T)$ to the zero temperature
equilibrium case. The top frame is for s-wave while the bottom is
d-wave. In each frame the short dashed curve is the approximate result at
$T=0$ derived above. We see that it compares well with the exact
result (solid curve). The other curves apply to $T/T_c=t=0.3$ (dotted),
0.5 (short-dashed), 0.7 (long-dashed) and 0.9 (dot-dashed). In this
case the $\Delta(n,T)/\Delta(0,0)$ curves do not cross and are
all constructed from BCS curves for the temperature dependence of the
gap. The temperature $T$ refers to the sample temperature before
the injection of excess quasiparticles $n$. The intersection of the
various curves with the vertical axis simply gives the 
temperature variation of the gap
in BCS. At finite $n$, the extra quasiparticles are
accommodated into the system by assuming
a higher temperature thermal distribution
$T^*$, with $T^*$ made sufficiently larger than $T$ to have
$n$ extra thermal quasiparticles. 

Note that in contrast to the $\mu^*$ model, the differences
between s- and d-wave are much less
pronounced at $T=0$. 
This reflects the fact that in a thermal distribution,
blocking effects are not an important consideration.
In fact now the gap in the
d-wave case terminates at a value of $n$ which is smaller than in the
s-wave case. This is opposite to what is found for the $\mu^*$ model.
Also the curves show no first order transition to the normal state which 
now occurs only when the gap is zero.

\begin{figure}[h]
\begin{picture}(250,200)
\leavevmode\centering\includegraphics{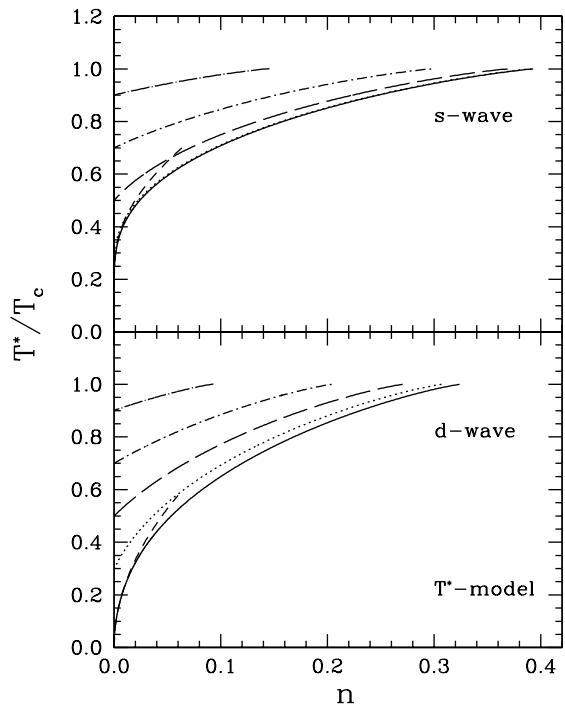}
\end{picture}
\vskip 60pt
\caption{The parameter $T^*/T_c$ versus $n$ for several temperatures
shown for the s-wave (top frame) and d-wave (bottom frame) gaps.
Curves are shown for
$T/T_c = t =0.01$ (solid curve), 0.3 (dotted), 0.5 (long-dashed),
0.7 (dot-short-dashed), 0.9 (dot-long-dashed). 
 The short-dashed curve is the small n limit (see Table II).
$T^*/T_c$ goes to $T/T_c$ as $n \to 0$, which forms the lower limit
on the curves with the upper limit being $T^*/T_c=1$ at which point,
$\Delta(T^*)$ would be zero.
}
\label{fig7}
\end{figure}

In Fig.~7 we show the value of $T^*$ as a function of the
nonequilibrium distribution $n$ for various values of 
$T$. The temperatures used are $T/T_c=
0.01$, 0.3, 0.5, 0.7, 0.9.
Note that the curve with the lowest sample temperature (solid curve)
at small $n$ agrees well with our analytic expressions for the same
quantity shown as the dashed lines. These follow from the
transcendental equation
$n=0.94\sqrt{T^*/ T_c} e^{-1.76 T_c/T^*}$
for s-wave and the explicit equation $T^*/T_c=2.36n^{1/2}$
for d-wave. These results are also entered in the final line
of Table~II.

\section{S-I-N tunneling junction}

Now we consider a specific application of our results to the 
case of a superconducting-insulator-normal metal 
tunneling junction.
Denote the current in a S-I-N junction with nonequilibrium
distribution on the superconducting side, described by the 
$\mu^*$ model, by $I^{SN}_{\mu^*}(V)$ where $V$ is the voltage
across the junction. It is given by a straightforward
modification of the usual tunneling formula\cite{parks} 
\begin{equation}
I^{SN}_{\mu^*}(V)=\int^\infty_{-\infty} d\epsilon \bar N_S(\epsilon)
[f(\epsilon-\mu^*)-f(\epsilon+V)]
\end{equation}
where $\bar N_S(\epsilon)$
is the normalized density of states given by
\begin{eqnarray}
\bar N_S(\epsilon) =\Re e \biggl\langle\frac{|\epsilon|}{\sqrt{\epsilon^2-\Delta^2_k}}
\biggr\rangle
\end{eqnarray}
with $<...>$ the average over angles as before.

We have seen in the previous section that the introduction of
a nonequilibrium $\mu^*$ 
modifies the gap but does not change its symmetry and Eqn.~(26)
still holds for the density of states in Eqn.~(25) although the new gap
amplitude is reduced by a factor of $(1-2n)$ and $(1-4\sqrt{2}n^{3/2}/3)$
for s- and d-wave, respectively, at zero temperature
and $n$ small. Besides the change in $\bar N(\epsilon)$ just described,
one of the thermal factors in Eqn.~(25) is also displaced by the new chemical
potential $\mu^*$. The structure of 
Eqn.~(25) makes it useful to separate 
these two factors, and it is convenient to rewrite $I^{SN}_{\mu^*}(V)$
in the form
\begin{eqnarray}
I^{SN}_{\mu^*}(V)&=&\int^\infty_{-\infty} d\epsilon \bar N_S(\epsilon)
[f(\epsilon-\mu^*)-f(\epsilon)]\nonumber\\&+&\int^\infty_{-\infty} d\epsilon
\bar N_S(\epsilon)[f(\epsilon)-f(\epsilon+V)]
\end{eqnarray}
The second term in Eqn.~(27) has the identical form that applies to
an ordinary S-I-N junction in equilibrium
at temperature $T$. We denote the current in this case by
${\cal I}(V)$
\begin{equation}
{\cal I}(V)\equiv \int^\infty_{-\infty} d\epsilon \bar N_S(\epsilon)
[f(\epsilon)-f(\epsilon+V)]
\end{equation}
where the gap amplitude defining $\bar N_S(\epsilon)$ is that
appropriate to the nonequilibrium superconductor. The first
term in Eqn.~(27) is simply a number, independent of voltage. Reference to
the defining Eqn.~(3) for $n$ shows
that this number is equal to $n\Delta(0)$.
Thus we find
\begin{eqnarray}
I^{SN}_{\mu^*}(V) &= {\cal I}(V)+n\Delta(0)  
\end{eqnarray}
We see from Eqn.~(29) that the current voltage characteristics
are modified in two ways by the nonequilibrium distribution. The
entire equivalent equilibrium distribution is shifted up by an
amount $n\Delta(0)$. This allows one to measure $n$
once the gap is known. Secondly, the ``equivalent equilibrium''
current voltage characteristics are those of an equilibrium junction
with the smaller nonequilibrium gap used instead of its
equilibrium value. This knowledge allows one to fully characterize
the non-equilibrium current voltage characteristic and to apply checks
to see how well the $\mu^*$ model works. For example, the derivative
of $I^{SN}_{\mu^*}(V)$ with $V$ at zero temperature simply gives 
\begin{equation}
\frac{dI^{SN}_{\mu^*}}{dV}=\bar N^S(V),
\end{equation}
the quasiparticle density of states with the nonequilibrium gap
but otherwise it is the same as for an equilibrium distribution.
For s-wave it will have an inverse square root singularity at $\Delta(n)$ 
and for 
d-wave it will go like  $\ln(8\Delta(n)/|\Delta(n)-\omega|)/\pi$
instead (see Abanov-Chubukov\cite{chubukov}).
In both cases, $\Delta(n)$ can be determined
from these singularities. Comparison
with its equilibrium value gives a measure of $n$ in both s- and d-wave.
Next it should be possible to check if this is consistent with the 
value of the chemical potential related to $n$ by $\mu^*/\Delta(0)
=1-2n$ and $\mu^*/\Delta(0)=\sqrt{2}n^{1/2}$,
respectively, for s- and d-wave, at $T=0$ and in the limit of small
$n$. The chemical potential is measured directly by noting that in
Eqn.~(27) for 
$V=-\mu^*$, first and second terms on the right-hand side are equal
but of opposite sign, giving a sum of 
zero. 
In Fig.~8, we show numerical results for 
$I^{SN}_{\mu^*}(V)$ versus $V$ at a low temperature $T/T_c=0.1$. The
top frame applies to s-wave while the bottom frame is for d-wave. It 
is verified that these curves obey the expected rules mentioned above.
For the s-wave case $\Delta(n)/\Delta(0)$ 
is set equal to 0.8 while for the d-wave case we have used
$\Delta(n)/\Delta(0) =0.9$ instead. Reference to Fig.~1 for s-wave and
to Fig.~3 for d-wave shows that these choices correspond to an
excess quasiparticle number of approximately 0.09 and 0.12,
respectively. The excess quasiparticle number is greater in the d-wave
case than in s-wave even though the gap is only reduced by 10\% as
compared with 20\% for s-wave.

\begin{figure}[h]
\begin{picture}(250,200)
\leavevmode\centering\includegraphics{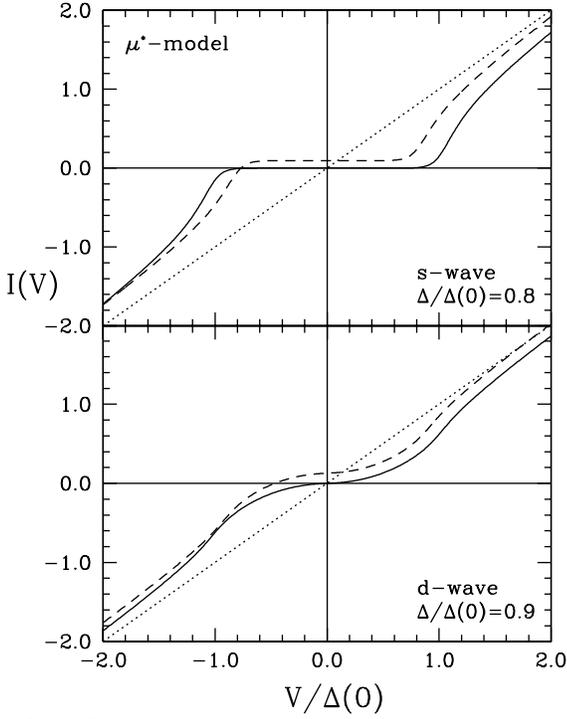}
\end{picture}
\vskip 60pt
\caption{SIN tunneling I-V characteristics
for $T=0$ shown for the $\mu^*$ model
with s-wave gap symmetry (upper frame) and d-wave 
(lower frame). The current $I$ is normalized by $N(0)$ and by
the maximum zero-temperature gap in the standard way and the
voltage $V$ is normalized to the maximum gap at $T=0$.
An excess quasiparticle density $n$ leads to a 
reduction in the gap by $\Delta(n)/\Delta(0)$ 
and a vertical shift by $n$ in the I-V characteristic. $I=0$ at $V=-\mu^*$.
The dotted curve is the normal state $n=0$,
the solid curve is superconducting state with $n=0$ and the
dashed curve is for a reduced gap $\Delta(n)/\Delta(0)=0.8$ in the
s-wave case and 0.9 in the d-wave case.}
\label{fig8}
\end{figure}

\section{Pump/probe optical measurements of $n(T)$}

In the following, we wish to discuss recent experimental pump/probe laser
experiments which have been used to infer information about the
excess quasiparticle density. In particular, we wish to address a claim
that these experiments provide evidence for s-wave pairing in the high $T_c$
cuprates. To address this issue, following Kabanov {\it et al.}\cite{kabanov},
we use the $T^*$ model. While we do not report explicitly on this here,
we have also examined these properties within the $\mu^*$-model
and have found similar results.

To calculate the excess quasiparticle density $n(T)$, we have
used Eqn.~(3) modified for the $T^*$ model via Eqn.~(4) such that
\begin{equation}
n(T) = \frac{1}{\Delta(0)}\Biggl<\int_0^\infty[f(E^*_k,T^*)
-f(E_k,T)]d\epsilon_k\Biggr>
\end{equation}
where $E_k^*=\sqrt{\epsilon_k^2+\Delta_k^2(T^*)}$ (the asterisk always
referring to quantities depending on $T^*$ instead of $T$)
and an average over the angle $\phi$ is done in the case of d-wave.

To evaluate $n(T)$ at a temperature $T$, it is necessary to know
$T^*$ and $\Delta(T^*)$ and this is determined from the amount of 
laser energy $E_I$ deposited in the system. In this work, the laser
energy will be assumed to go into both electron and phonon
systems 
\begin{equation}
E_I=\Delta E_{electron}+\Delta E_{phonon}.
\end{equation}
To begin with, however, we examine the case where the energy is assumed to
go only into the electronic system: the quasiparticles and a modification
of the superconducting condensate due to a $\Delta(T^*)$. The energy
going into the quasiparticles relative to the reference nonequilibrium 
state at temperature $T$ is
\begin{equation}
\Delta E_{qp} = 4N(0)\Biggl<\int_0^\infty[E_k^*f(E^*_k,T^*)
-E_kf(E_k,T)]d\epsilon_k\Biggr>.
\end{equation}
Kabanov {\it et al.}\cite{kabanov} treated this piece as 
$[n(T^*)-n(T)](\Delta(T)+k_BT/2)$ 
which is not completely correct near $T_c$. 

\begin{figure}[h]
\begin{picture}(250,200)
\leavevmode\centering\includegraphics{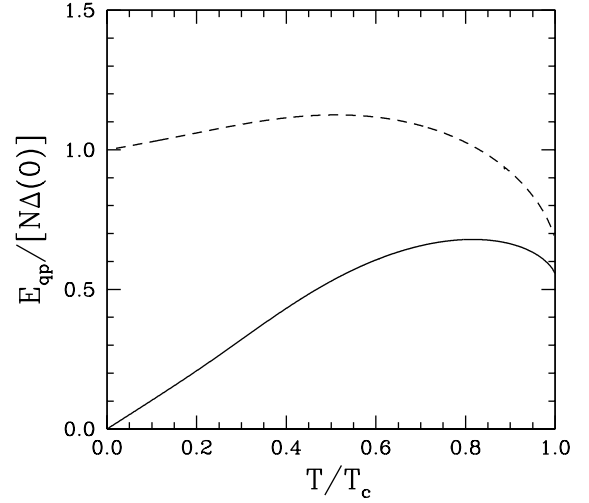}
\end{picture}
\caption{The average quasiparticle energy per particle
normalized to the maximum zero temperature equilibrium gap,
$E_{qp}/N\Delta(0)$, versus  $T/T_c$ for s-wave
(dashed curve) and d-wave (solid curve). The $T=0$ value is
set by the lowest available energy state in the quasiparticle density of
states, whereas near $T_c$, the energy scale is set by $k_BT$. 
}
\label{fig9}
\end{figure}

We find
that the average quasiparticle energy per particle calculated as 
\begin{equation}
\frac{E_{qp}}{N} = \frac{\int_0^\infty E_kf(E_k,T)d\epsilon_k}
{\int_0^\infty f(E_k,T)d\epsilon_k}
\end{equation}
gives a constant equal to $\Delta(0)$ at zero temperature for s-wave,
since excitations can only exist at the gap edge because the
density of states is zero below this energy. This behaviour is
seen in Fig.~9 for the dashed curve which gives
Eqn.~(34) normalized to $\Delta(0)$. As $T$ increases, the energy
per particle increases slightly and then decreases near $T_c$ to
a value of $\pi^2k_B T_c/12\ln(2)\Delta(0) = 1.19(k_BT_c/\Delta(0))
\simeq 0.67$ as now the gap in
the quasiparticle density of states has shrunk to zero and the energy of
the quasiparticles is controlled by $k_BT$ which is less than $\Delta(0)$.
Similar physics is found for a d-wave order parameter with the essential
difference that excitations can now occur at zero energy 
and therefore the average quasiparticle
energy per particle starts from zero at $T=0$ and rises linearly
reflecting the linear increase in energy of the density of states.
It can be shown analytically that 
$E_{qp}/N\Delta(0) \simeq 1.03 T/T_c$ for
$T\ll\Delta(0)$, the regime where a nodal approximation is 
valid.
At $T_c$, the quasiparticle energy per particle is once
again controlled by $k_BT$ and so the limiting number is given by the
same formula as above but with the BCS d-wave gap ratio of $\Delta(0)/k_BT_c
= 2.14$ instead of 1.76 for s-wave. The number at $T_c$ is 
approximately 0.55
These results are shown in Fig.~9 (solid curve) 
and we will refer back to them at
a later point.

The reaction of the condensate is simply given as
\begin{eqnarray}
\Delta E_{cond} &=& 
2N(0)
\Biggl<\int_0^\infty[E_k-E_k^*+\frac{\Delta^{*2}_k}{2E_k^*}(1-2f(E^*_k,T^*))
\nonumber\\&-&\frac{\Delta^{2}_k}{2E_k}(1-2f(E_k,T))]d\epsilon_k\Biggr>.
\end{eqnarray}
This term reflects the fact that the presence of excess quasiparticles 
causes a readjustment to the superconducting condensate through a change in
the gap
$\Delta^*\equiv\Delta(T^*)$. This term was not
included by Parker\cite{parker}  and neither was it included in the work of
Kabanov {\it et al.}\cite{kabanov}.

\begin{figure}[h]
\begin{picture}(250,200)
\leavevmode\centering\includegraphics{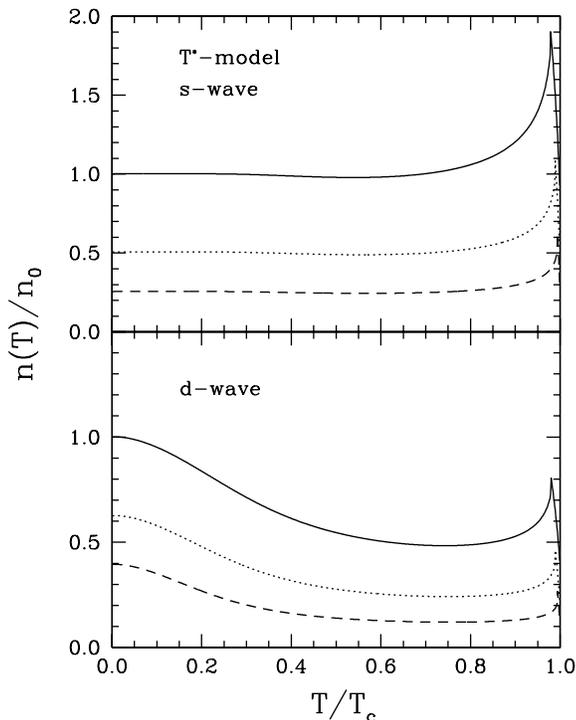}
\end{picture}
\vskip 60pt
\caption{The excess quasiparticle fraction, $n(T)/n_0$
as a function of $T/T_c$. These curves are calculated in
 the $T^*$ model for
fixed laser energy, where the energy is
 assumed to go only into the quasiparticles. The upper frame is
for a BCS s-wave gap and the bottom frame is for d-wave.
Curves are shown for fixed laser energy $E_I$ (in units of the
condensation energy) of 0.2 (solid), 0.1 (dotted), and 0.05 (dashed).
Here $n_0$ is $n(T=0)$ for the case of $E_I=0.2$.
}
\label{fig10}
\end{figure}

Our procedure was to fix the laser energy $E_I$ and determine
the $T^*$ and $\Delta(T^*)$ which gave $E_I=\Delta E_{qp}+\Delta E_{cond}$.
For our purposes, we used the BCS temperature dependence of the gap,
calculated numerically, for both $\Delta(T^*)$ versus $T^*$ and
$\Delta(T)$ versus $T$ with no approximate form.
Our results for both s-wave and d-wave gap symmetry are shown in 
Fig.~10 for a variety of $E_I$, which is normalized to the zero 
temperature condensation energy in the equilibrium state. 
Note that the curves shown here are normalized to $n_0\equiv n(T=0)$
for the $E_I=0.2$ case, rather than the $n(0)$ associated with each $E_I$,
in order to show the overall relative reduction as $E_I$ is reduced.
For s-wave, the 
curves are relatively flat albeit with some small
depression followed by  a sharp upturn near $T_c$ and then  by a drop.
The peak occurs when $T^*/T_c= 1$, at which point we assume 
that the nonequilibrium state has been forced to become a normal metal
at an effective temperature $T^*$ and it is measured relative to the
equilibrium superconducting state which would exist at temperature
$T$. Therefore, $\Delta E_{electron} = \Delta E_{electron}
(\Delta^*=0, T^*, \Delta(T), T)$ with $T^*$ being
fixed by $E_I$. Likewise, $n(T)=n(\Delta^*=0,T^*,\Delta(T),T)$.
The behaviour of the s-wave curve largely mimics the inverse of the
curve for the energy per quasiparticle.
At low temperature, the 
number of excess quasiparticles is relatively constant with a
slight decrease as the temperature is raised, reflecting the fact that
the energy per quasiparticle is increasing slightly and so fewer quasiparticles
can be created at fixed energy. At high temperature near $T_c$, the
energy per quasiparticle is decreasing and so more quasiparticles can
be created for fixed energy and one finds that $n(T)/n(0)$ shows an
upturn in response to this. Likewise, the d-wave curve for $n(T)/n(0)$
 can be understood from the behaviour of the $E_{qp}/N$ curve, with
the $n(T)/n(0)$ decaying dramatically as $T$ increases reflecting that
it is costing on average more energy to create a quasiparticle. 
We can easily show for the d-wave case that 
\begin{equation}
\frac{n(T)}{n(0)} = \biggl(1+\frac{2.9t^3}{E_I}\biggr)^{2/3}
-\biggl(\frac{2.9t^3}{E_I}\biggr)^{2/3}
\end{equation}
for small reduced temperature $t=T/T_c$ with $T\ll\Delta(0)$. 
For $t\ll E_I$, $n(T)/n(0)\simeq 1-(2.9/E_I)^{2/3}t^2$.
For $t\gg  E_I$ but still with $T\ll\Delta(0)$,
 $n(T)/n(0)\simeq (2/3t)(E_I/2.9)^{1/3}$ (inverse $t$ law).
Our numerical results conform to these limits. Also note that
$n(0)=0.18(E_I/2.9)^{2/3}$  so that $n(T)$ unnormalized to $n(0)$
will go like $E_I$ in the region where the $1/t$ law applies.
Once again as the energy scale reverts to $k_BT$ near $T_c$ the slight
upturn in $n(T)/n(0)$ is reflecting the smaller energy required to
create the quasiparticles. Kabanov {\it et al.}\cite{kabanov}
 do not find this result due to their
approximations and the details of their curve would differ as they
have only included an approximate linear form of the d-wave
quasiparticle density of states rather than the full form with
temperature dependence as is done here. In fact, if their data did not
go so low in temperature and given that $n(0)$ is not known experimentally, 
the flatness of the d-wave curve with a slight upturn near $T_c$ placed
on an arbitrary scale, would probably 
make as viable a comparison with their data
as the s-wave case. However, we note that they do show data at lower 
temperature and so this interpretation does not hold,
also the $2/3$ dependence on $E_I$ at $T=0$ is not verified.
On the face of things, it may appear that their data agree best with 
s-wave. However, we argue, as they did, that it is necessary to
include  phonons in this picture.

In their analysis, to obtain agreement with their data, Kabanov
{\it et al.}\cite{kabanov}
 did include the fact that some of the laser energy would be 
distributed to phonons in the system.
In this case, we partition the laser energy with the phonons as
well:
\begin{equation}
E_I = \Delta E_{qp}+\Delta E_{cond} +\Delta E_{phonon}
\end{equation}
The phonon piece is calculated assuming that only phonons with energy
$\hbar\omega$
above $2\Delta$ can be considered to be out of thermal equilibrium
with the lattice and therefore at a temperature $T^*$. 
It is in this way that a bottleneck at $2\Delta$ is introduced
into the model, which then deviates from pure heating. Such was the
same consideration of Kabanov {\it et al.}\cite{kabanov}. 
As such we calculate the
amount of energy going into the phonon system as:
\begin{equation}
\Delta E_{phonon}=\int_{2\Delta}^\infty \omega F(\omega)
[n(\omega,T^*)-n(w,T)]d\omega,
\end{equation}
where the usual Bose-Einstein factor
$n(\omega,T)=1/({\rm exp}(\hbar\omega/k_BT)-1)$ and $F(\omega)$ is the phonon
frequency distribution function measured by Renker {\it et al.}\cite{renker} from
neutron scattering experiments on YBCO. 
In our calculation, we effectively fix $N(0)$ to get the correct ratio of 
phononic specific heat at $T_c$ relative
to the electronic part via comparison with the specific heat data
of Loram {\it et al.}\cite{loram}. This is to ensure that,
the phononic and electronic portions are balanced in accordance with 
experiment. 
As the phonon energy increases typically as $T^4$, one sees that this term, as 
long as $T^*>T$, will take more and more of the fixed laser energy
away from the electronic system and hence, there are fewer excess 
quasiparticles that can be created at higher temperature and the 
curve for $n(T)/n(0)$ 
must go down. This is illustrated in Fig.~11, where
curves decay rapidly as $T$ increases and are further reduced for lower
$E_I$. Also shown on the same figure are the experimental results of
Kabanov 
{\it et al.}\cite{kabanov} (solid squares). We
conclude that their data does not support an interpretation of s-wave
gap symmetry in the high $T_c$ cuprates. Nor does it agree with
d-wave (Fig. 10, bottom frame).

\begin{figure}[h]
\begin{picture}(250,200)
\leavevmode\centering\includegraphics{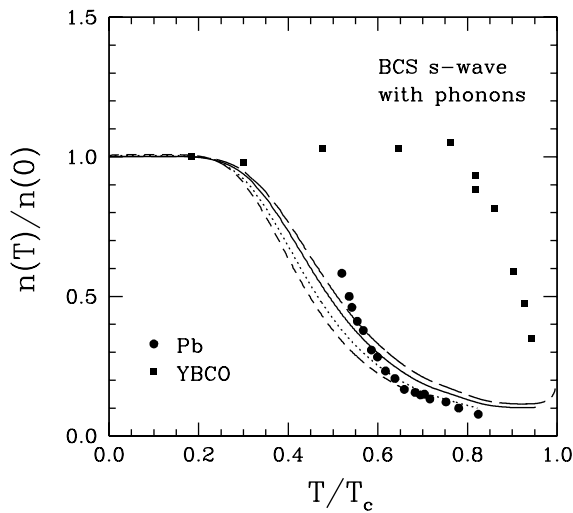}
\end{picture}
\caption{The excess quasiparticle fraction
$n(T)/n(0)$ 
using a $T^*$ model in s-wave
including phonons for parameters appropriate to YBCO.
For Pb parameters the curves are very similar (and not shown). 
The curves are for fixed laser energy (in units of the
condensation energy) of 0.3 (long-dashed),
0.2 (solid), 0.1 (dotted), and 0.05 (short-dashed).
The Pb data of Carr {\it et al.}\cite{carr}, 
reproduced here as the black dots, shows
a suppression with increasing
$T$ in keeping with our s-wave results with phonons.
The YBCO data of Kabanov {\it et al.}\cite{kabanov}
are shown as the solid squares and disagree with the theory.
}
\label{fig11}
\end{figure}

We have also done this calculation with the Pb phonon spectrum\cite{pba2f}
 (adjusted to the specific heat in Pb and using a BCS gap ratio) and we find
similar curves to those shown here.
The excess quasiparticle
density in Pb has been measured by Carr {\it et al.}\cite{carr} and compared
successfully to rate equation calculations\cite{kaplan} used for
determining the nonequilibrium distribution.
We show the Pb data (solid circles) on our
curves to emphasize that Pb, as an s-wave superconductor, does follow
the trend of showing a suppressed excess quasiparticle density as 
the temperature increases and agrees well with our calculations. 
This comparison also serves to show that 
our simple procedure of introducing the bottleneck at $2\Delta$
and sharing the laser energy between phonons and electrons agrees
qualitatively and even semiquantitatively with the  more
sophisticated and accurate
rate equation calculations used by Carr {\it et al.}\cite{carr}
and validates our simpler method.

Here, we have done the calculation using a
BCS gap ratio of 3.53. A full strong coupling Eliashberg calculation\cite{carbotte}
would have to be done to include a larger ratio, as from our experience,
simply inserting a larger ratio in a BCS calculation can give 
incorrect, and therefore misleading, results. Aside from the
inherent complexity of such a calculation, we would need to commit to
some specific mechanism since 
phonons are not believed to
be the source of the high $T_c$.
But there is no consensus on mechanism. 
To fit 
experiment, however, Kabanov {\it et al.}\cite{kabanov}
phenomenologically increase the value of the ratio
$2\Delta(0)/k_BT_c$ to about 9. 
There is, however, no rigorous justification
for such a procedure and this is our main objection to such a
fit. To increase the gap ratio, it is necessary to increase the
ratio of $T_c/\omega_{\ln}$ in Eliashberg theory\cite{carbotte}, where
$\omega_{\ln}$ is a particular moment of the electron-phonon
spectral function which gives the appropriate measure of the
average phonon energy involved. When this
is done, damping effects, entirely left out of BCS, become dominant
and superconducting properties acquire behaviours that
are qualitatively different from straightforward extrapolations
of BCS behaviour (see, for instance, many properties calculated in
Ref.~\cite{carbotte} in the limit of large $T_c/\omega_{\ln}$ ratio).
 For YBCO, the gap ratio is closer to
5\cite{YBCOgap} and is certainly not 9. Further, for a gap ratio
ratio of 9-10, the cutoff of $2\Delta$ applied to the phonons
falls at 70-80 meV which is at the very top of the measured phonon 
spectrum\cite{renker}. This large value of the cutoff
has the effect of greatly reducing the ability
of the phonons to share in the laser energy and this partially accounts for
why the curve for  $n(T)$ in this case stays flat to much higher
temperature than for the BCS curve.

\section{Conclusions}

In summary, we have examined the differences between an s-wave order
parameter versus a d-wave in a nonequilibrium superconductor
using two prominent models in the literature,
the  $T^*$ model of Parker\cite{parker} and the $\mu^*$ model
of Owen and Scalapino\cite{owen}. While these models may
be considered to be somewhat crude, they have the virtue of being
simple, and accessible in terms of both calculation and
physical intuition. As a result, one finds them still being used by
experimental groups to aid in the interpretation of their data.
With the advent of high $T_c$ cuprates and the deeper examination 
of the issue of order parameter symmetry, a re-examination of
these models allows for the prediction of different power law
dependences on excess quasiparticle density expected 
between s- and d-wave gaps. Tables I and II summarize such predictions.
These predictions are grounded in interesting physics such as the blocking
of states and how the condensate readjusts and as such, they should remain
relevant even within more complicated models. It is hoped that the
simplicity of our results may inspire further experimental and
theoretical efforts to examine nonequilibrium phenomena in the presence of
an order parameter with nodes.

In addition, we remind the reader of the past discussions of an inhomogeneous 
state in the $\mu^*$-model and provide the prediction that the d-wave
state will not be unstable to such a state for the most part. In fact,
the d-wave state may form a more stable and intuitively interesting
state in which to probe nonequilibrium superconductivity.

In our work we have specifically addressed two experiments. 
SIN tunneling has always 
been a powerful probe of s-wave superconductors and with our work within the
$\mu^*$-model, we show how one may use this experiment to measure
the model parameters of $\mu^*$, $n$ and $\Delta(n)$ in order to
test the predictions of the theory, both for s-wave and d-wave cases.
The modern use of STM may provide a more attractive avenue for
investigating this issue in the face of inhomogeneity where 
the local density of states may vary with position within the same
material.
 
Finally, recent pump/probe experiments in YBCO which have been interpreted 
as providing support for s-wave gap symmetry\cite{kabanov} are reconsidered.
Within a BCS description of the superconducting state and a $T^*$ model
for the nonequilibrium distribution, our calculations, including
phonons, do not produce an excess
quasiparticle distribution $n(T)$ which is nearly constant in
temperature with a peak near $T_c$. Rather a quick decay with increasing $T$ is
found as more of the laser energy is taken up by the phonon system.
When the explicit case of Pb is considered rather than YBCO, the same
rapidly decaying characteristic is found and this is in good
agreement with the recent data of Carr {\it et al.}\cite{carr} in
this classic s-wave superconductor. Our final conclusion is that 
present pump/probe experiments in YBCO cannot be accounted 
for by either s- or d-wave gap symmetry and it may
be necessary to re-examine the interpretation of the data in terms of the
excess quasiparticle density. In this regard, a next step might
be to calculate the optical conductivity itself in the nonequilibrium
state so as to make a more direct contact with what is measured.

\begin{acknowledgments}
EJN acknowledges funding from NSERC, Research Corporation, the
Government of Ontario
(Premier's Research Excellence Award), and the University of Guelph.
JPC acknowledges support from NSERC and the CIAR. We 
thank Larry Carr, 
David Tanner, David Smith, Ilya Vekhter,
and Bernie Nickel for helpful discussions.
\end{acknowledgments}

\end{document}